\documentclass[12pt]{article}
\newcommand{\beq}{\begin{equation}}
\newcommand{\eeq}{\end{equation}}
\newcommand{\beqa}{\begin{eqnarray}}
\newcommand{\eeqa}{\end{eqnarray}}
\newcommand{\beqar}{\begin{eqnarray*}}
\newcommand{\eeqar}{\end{eqnarray*}}
\newcommand{\hph}[1]{{\hphantom{#1}}}

\newcommand{\ga}{\gamma}

\newcommand{\Om}{\Omega}

\newcommand{\eg}{{\it e.g.,}\ }
\newcommand{\ie}{{\it i.e.,}\ }
\newcommand{\labell}[1]{\label{#1}} 
\newcommand{\reef}[1]{(\ref{#1})}
\newcommand\prt{\partial}

\newcommand\cR{{\cal R}}

\newcommand\cF{{\cal F}}

\newcommand\cM{{\cal M}}

\newcommand\cB{{\cal B}}

\newcommand\tG{{\widetilde G}}

\newcommand\Tr{{\rm Tr}}

\usepackage{color}
\begin{document}

\begin{titlepage}

\begin{center}



\vskip 2 cm
{\Large \bf Higher derivative corrections to DBI action at  $ \alpha'^2$  order
 }\\
\vspace*{1cm}

 Komeil Babaei Velni$^{1}$\footnote[1]{babaeivelni@guilan.ac.ir} and   Ali Jalali$^{2}$\footnote[2]{ali.jalali@stumail.um.ac.ir}\\
\vspace*{1cm}
$^{1}${Department of Physics, University of Guilan,\\ P.O. Box 41335-1914, Rasht, Iran}
\\
\vspace{0.5cm}
$^{2}${Department of Physics, Ferdowsi University of Mashhad,\\ P.O. Box 1436, Mashhad, Iran}
\\
\vspace{2cm}

\end{center}

\vskip 0.5 cm

\begin{abstract}
\baselineskip=18pt
We use the compatibility of  D-brane  action  with linear off-shell T-duality and linear on-shell S-duality  as  guiding principles to find all world volume couplings of one  massless closed    and three  massless open strings   at order  $\alpha'^2$ in type II superstring theories in flat space-time.
\end{abstract}
Keywords:   T-duality; S-duality; Higher-derivative Couplings 
\end{titlepage}

\section{Introduction and Results}

D-branes are non-perturbative objects in superstring theory which play a central role in exploring  different  aspects of the theory; from  statistical computation of black hole entropy \cite{Strominger:1996sh} to realization of the AdS/CFT correspondence \cite{Maldacena:1997re}  or appearance of noncommutative geometry in string theory \cite{Seiberg:1999vs}. However, these objects can be studied by using a perturbative method  thanks to their description in terms of open strings with Dirichlet boundary conditions \cite{Green:1996um}.

Much of the importance of D$_p$-branes stems from the fact that they provide a remarkable way of introducing nonabelian gauge symmetries in string theory. Nonabelian gauge fields naturally appear confined to the world volume of multiple coincident D$_p$-branes \cite{Taylor:1999pr,Myers:1999ps}.

The  low energy effective field theory of  D$_p$-branes  in type II superstring theories consists of the  Dirac-Born-Infeld (DBI) \cite{Bachas:1995kx} and the  Chern-Simons (CS) actions \cite{Polchinski:1995mt}, \ie
\beqa
S_p&=&-T_p\int d^{p+1}x\,e^{-\phi}\sqrt{-\det\left(P[g+B]_{ab} +F_{ab} \right)}+T_{p}\int e^FP[e^{B}C] \labell{DBI}
\eeqa
where $P[\cdots]$ is the pull-back operator which projects the spacetime tensors to the world volume, \eg 
 $P[g]_{ab}={\prt X^{\mu}}/{\prt\sigma^a} {\prt X^{\nu}}/{\prt\sigma^b}g_{\mu\nu}=\tG_{ab}$. The dependence of the closed string fields on the transverse coordinates appears in the action via  the Taylor expansion \cite{Garousi:1998fg}.  In the literature, there is a factor of $2\pi\alpha'$ in front of gauge field strength $F_{ab}$ while here, we normalize the gauge field to  absorb this factor.

The curvature, the second fundamental form and the dilaton corrections  to the DBI action at order $\alpha'^2$ in the string frame are reported  in \cite{Bachas:1999um,Garousi:2011fc,Garousi:2009dj} as\footnote{Our index convention is that the Greek letters  $(\mu,\nu,\cdots)$ are  the indices of the space-time coordinates, the Latin letters $(a,d,c,\cdots)$ are the world-volume indices and the letters $(i,j,k,\cdots)$ are the normal bundle indices. The Killing index in the reduction of 10-dimensional spacetime to 9-dimensional spacetime is $y$.}
\beqa 
S_p^{DBI} &\supset&-\frac{\pi^2\alpha'^2T_{p}}{48}\int d^{p+1}x\,e^{-\Phi}\sqrt{-\tG}\bigg[(R_T)_{abcd}(R_T)^{abcd}-2(\cR_T)_{ab}(\cR_T)^{ab}\nonumber\\
&&\qquad\qquad\qquad\qquad\qquad\qquad-(R_N)_{abij}(R_N)^{abij}+2\bar{\cR}_{ij}\bar{\cR}^{ij}\bigg]\labell{RTN}
\eeqa
where  
 the  curvatures $(R_T)_{abcd}$ and $(R_N)^{abij}$  are related to the projections of  the bulk Riemann curvatures into world volume and transverse spaces via the Gauss-Codazzi equations, \ie
\beqa
(R_T)_{abcd}&=&R_{abcd}+\delta_{ij}(\Omega_{\ ac}{}^i\Omega_{\ bd}{}^j-\Omega_{\ ad}{}^i\Omega_{\ bc}{}^j)\nonumber\\
(R_N)_{ab}{}^{ ij}&=&R_{ab}{}^{ij}+g^{cd}(\Omega_{\ ac}{}^i\Omega_{\ bd}{}^j-\Omega_{\ ac}{}^j\Omega_{\ bd}{}^i)\labell{RTRN}
\eeqa
In addition, the   curvatures $(\cR_T)_{ab}$ and $\bar{\cR}_{ij}$  are related to the   Riemann curvatures, the   second fundamental form and the dilaton via the following relations:
\beqa
(\cR_T)_{ab}\!\!\!&=&g^{cd}(R_T)_{cadb}+\prt_a\prt_b\Phi\nonumber\\
 (\bar{\cR})^{ij}&=&g^{ab}R_a{}^{i}{}_b{}^{j}+g^{ab}g^{cd}(\Omega_{\ ac}{}^i\Omega_{\ bd}{}^j-\Omega_{\ ab}{}^i\Omega_{\ cd}{}^j)+\prt_i\prt_j\Phi\labell{Rij}
\eeqa
in which the world volume indices are raised by the inverse of the pull-back metric.
In static gauge, the second fundamental form   includes the second derivative of the transverse scalar fields, \ie $\Omega_{ab}{}^i=\prt_a\prt_b\phi^i-\tilde{\Gamma}_{ab}{}^c\prt_c\phi^i+\Gamma_{ab}{}^i$. Thus, the  action \reef{RTN} includes the couplings of one graviton or dilaton and two scalar fields. All other couplings between one NSNS and two NS fields at order $\alpha'^2$ have been found in \cite{Jalali:2015xca} by requiring  \reef{RTN} to be invariant under linear T-duality and also
 the couplings   to be consistent  with the corresponding S-matrix elements. The couplings in the string frame are \cite{Jalali:2015xca} 
 \beqa
S_p^{DBI} &\!\!\!\!\!\supset\!\!\!\!\!&-\frac{\pi^2\alpha'^2T_{p}}{12}\int d^{p+1}x\,e^{-\Phi}\sqrt{-\tG}\bigg[ \cR_{bd}\big(\prt_{a}F{^{ab}}\prt_{c}F^{cd}-\prt_{a}F_{c}{}^{d}\prt^{c}F^{ab}\big)+
\frac{1}{2}R_{bdce}\prt^{c}F^{ab}\prt^{e}F_{a}{}^{d}\nonumber\\
&& \qquad\qquad\qquad\qquad+\frac{1}{4}\cR_{d}{}^{d}\big(\prt_{a}F^{ab}\prt_{c}F_{b}{}^{c}+\prt_{b}F_a{}^{c}\prt_{c}F{}^{ab}\big)+\Om_{a}{}^{ai}\prt_{d}H_{c}{}^{d}{}{}_{i}\prt_{b}F^{bc}\nonumber\\
&& \qquad\qquad\qquad\qquad-\Om^{bai}\bigg(\prt_{b}F_{a}{}^{c}\prt_{d}H_{c}{}^{d}{}_{i}
+\prt^{d}F_{a}{}^{c}\prt_{i}H_{bcd}-\frac{1}{2}
\prt^{d}F_{a}{}^{c}\prt_{c}H_{bdi}\bigg)\bigg]\labell{DBIj}
\eeqa
 where 
  $\cR_a{}^a$ is given by
\beqa
\cR_{a}{}^{a}&=& R^{ab}{}_{ab}+2\prt^a\prt_a\Phi\labell{del}
\eeqa
which is invariant under linear T-duality.  
In \reef{DBIj}, the   Riemann curvatures and the 
field strength $ H = d B$,   are  the pull-back of the spacetime   curvature and B-field onto world volume  and  transverse spaces respectively.
 For example
$$R_{abcd}=\prt_aX^\alpha\prt_bX^\beta\prt_cX^\mu\prt_dX^\nu R_{\alpha\beta\mu\nu}$$
or
$$\prt_a H_{bci}\Omega_{de}{}^i=\bot^{\alpha\beta}\prt_aX^\mu\prt_bX^\nu\prt_cX^{\rho}\prt_dX^\sigma\prt_eX^\zeta \prt_\mu H_{\nu\rho\alpha}\Omega_{\sigma\zeta\beta}\,,$$
in which $\prt_aX^\mu$ is the pull-back operator into the world volume while $\bot^{\alpha\beta}$ is the pull-back operator into the transverse space \cite{Corley:2001hg}  \ie
$$\bot^{\alpha\beta}=G^{\alpha\beta}-\tilde{G}^{\alpha\beta}\,,\qquad\tilde{G}^{\alpha\beta}=\frac{\prt X^\alpha}{\prt\sigma^a}\frac{\prt X^\beta}{\prt\sigma^b}\tilde{G}^{ab}\,.$$
In the last equation, $G^{\alpha\beta}$ is the first fundamental form and $\tG^{ab}$ is the inverse of the pull-back metric. In the static gauge, \ie $X^a=\sigma^a$ and $X^i=\phi^i$, the components of the projection operator $\bot^{\alpha\beta}$ become $\bot^{ab}=0$, $\bot^{ai}=-\prt^a\phi^i$ and $\bot^{ij}=\eta^{ij}$ to the linear order of transverse scalar field in which we are interested.
So one can extend the couplings   \reef{DBIj} from one NSNS closed string  and two open string couplings to the one NSNS and three open NS string. These couplings  are called  as $S_{p.l}$\,.

On the other hand  the four open string corrections to the DBI action at $\alpha'^2$ order have been found in \cite{Garousi:2015mdg}. These couplings, that appear in three families, include the couplings of four scalar fields, four gauge fields and two scalars and two gage fields. The four scalar field couplings are \cite{Garousi:2015mdg}:
\beqa
S_{\phi\phi\phi\phi}&=&-\frac{\pi^2\alpha'^2T_p}{48}\int d^{p+1}x\sqrt{-\tG}\bigg[  4 \Omega ^a{}^b{}^i \Omega _a{}_b{}^j \Omega ^c{}^d{}_i \Omega _c{}_d{}_j-4 \Omega ^a{}^b{}^i \Omega _a{}^c{}_i \Omega _b{}^d{}^j \Omega _c{}_d{}_j\nonumber\\
&&+4 \Omega ^a{}_a{}^i \Omega ^b{}^c{}_i \Omega _b{}^d{}^j \Omega _c{}_d{}_j -6 \Omega ^a{}_a{}^i \Omega ^b{}_b{}^j \Omega ^c{}^d{}_i \Omega _c{}_d{}_j+2 \Omega ^a{}_a{}^i \Omega ^b{}_b{}_i \Omega ^c{}_c{}^j \Omega ^d{}_d{}_j \bigg]\labell{fourph} 
\eeqa
In the above effective action, at the order of four fields, in  addition to the four scalar couplings there are couplings between one graviton and three scalar fields, two gravitons  and two scalar fields, three gravitons and one   scalar field and four gravitons. In this paper we are interested in the couplings between  one graviton and three scalars of Eq. \reef{fourph}. As discussed in  \cite{Garousi:2015mdg}, the consistency of   couplings in  \reef{fourph} with the T-duality and S-duality can be used to determine the four gauge field couplings as follow:
\beqa
S_{ffff}&\!\!\!\!\!=\!\!\!\!\!&-\frac{\pi^2\alpha'^2T_p}{48}\int d^{p+1}x\sqrt{\tG}\bigg[ 2 \prt^a\cF^b {}^c \prt_b\cF _a {}^d \prt_c\cF^e {}^f \prt_d\cF _e {} _f  \labell{fourf}\\&&- 
 \frac{1}{2} \prt_a\cF _b {} _c \prt^a\cF^b {}^c \prt_d\cF _e {} _f \prt^d\cF^e {}^
   f - 2 \prt^a\cF_a {}^b \prt^c\cF_b {} _c \prt_d\cF _e {} _f \prt^d\cF^e {}^
   f - 4 \prt^a\cF^b {}^c \prt_b\cF _c {}^d \prt_d\cF^e {}^
    f \prt_e\cF _a {} _f\nonumber\\&& + 
 2 \prt_a\cF^d {}^e \prt^a\cF^b {}^c \prt_f\cF _c {} _e \prt^f \cF_b {} _d + 
 2 \prt^a\cF_a {}^b \prt^c\cF_b {}^d \prt_d\cF _c {}^e \prt^f \cF_e {} _f + 
 6 \prt^a\cF_a {}^b \prt_c\cF^e {}^f \prt^c\cF_b {}^d \prt_d\cF _e {} _f\bigg]\nonumber
\eeqa
in which $\cF_{ab}=F_{ab}+B_{ab}$ is the gauge invariant combination of B-filed and gauge field. 
It should be pointed out that due to  the definition of $\cF_{ab}$, the couplings in Eq.  \reef{fourf}  also include two B-field and two gauge field couplings, three B-field and one gauge field couplings and four B-field couplings that  we do not consider  in this work.
Moreover, the couplings of two gauge fields and two scalar fields appear in the following effective action \cite{Garousi:2015mdg}:
\beqa
S_{\phi\phi ff}&\!\!\!\!\!=\!\!\!\!\!&-\frac{\pi^2\alpha'^2T_p}{48}\int d^{p+1}x\sqrt{-\tG}\bigg[ 
 4 \Omega_c {} _d {}^i \Omega^
   e {} _e {} _i \prt_a\cF _b {}^d \prt^a\cF^b {}^c -8 \Omega_c {}^e {}^
   i \Omega_d {} _e {} _i \prt_a\cF _b {}^d \prt^a\cF^b {}^c\nonumber\\&&+ 
 12 \Omega_c {}^e {}^
   i \Omega_d {} _e {} _i \prt^a\cF^b {}^c \prt_b\cF _a {}^d - 
 4 \Omega_c {} _d {}^i \Omega^
   e {} _e {} _i \prt^a\cF^b {}^c \prt_b\cF _a {}^d - 
 4 \Omega^d {} _d {}^i \Omega^
   e {} _e {} _i \prt^a\cF_a {}^b \prt^c\cF_b {} _c \nonumber\\&&- 
 4 \Omega_c {}^e {}^
   i \Omega_d {} _e {} _i \prt^a\cF_a {}^b \prt^c\cF_b {}^d + 
 12 \Omega_c {} _d {}^i \Omega^
   e {} _e {} _i \prt^a\cF_a {}^b \prt^c\cF_b {}^d + 
 8 \Omega_a {} _c {}^
   i \Omega_d {} _e {} _i \prt^a\cF^b {}^c \prt^d\cF_b {}^e \bigg]\labell{twoftwoph} 
\eeqa
 Through the $\cF_{ab}$ definition the couplings in \reef{fourf}  include either one graviton, one scalar and two gauge fields or one B-filed, one gauge field and two scalar field couplings. The couplings in \reef{twoftwoph}, in addition to the mentioned ones, include the couplings of two closed strings and two open strings, three closed strings and one open string and four closed strings that we do not consider here.
 
In this paper, we use the compatibility of couplings in   \reef{DBIj}, \reef{fourph}, \reef{fourf} and \reef{twoftwoph} and all other couplings of one NSNS closed string and three NS open strings
with off-shell linear T-duality and on-shell linear S-duality to find all couplings of one NSNS and three open string couplings at $\alpha'^2$ order. 
We  write all contractions of one  NSNS and three NS fields at order $\alpha'^2$
 with unknown coefficients. By imposing consistency of the couplings with  linear T-duality and linear S-duality, one can fix the coefficients. We find the graviton couplings as:
\beqa
S_{h\phi\phi\phi}&\!\!\!\!\!=\!\!\!\!\!&-\frac{\pi^2\alpha'^2T_p}{48}\int d^{p+1}x\sqrt{-\tG}\,F_{ab}
\bigg[R_{d}{}^{e}{}_{ei}\Om^{c}{}_{a}{}^{i}\prt_{b}F_{c}{}^{d}-2R_{d}{}^{e}{}_{ei}\Omega_{c}{}^{ci}\prt_{b}F_{a}{}^{d}\nonumber\\
&&-R_{cide}\Omega^{c}{}_{a}{}^{i}\prt_{b}F^{de}+R_{d}{}^{e}{}_{ei}\Omega^{dci}\prt_{c}F_{ab}+R_{bedi}\Omega^{dci}\prt_{c}F_{a}{}^{e}
-3R_{bide}\Omega^{dci}\prt_{c}F_{a}{}^{e}\nonumber
\\&&-2R_{d}{}^{e}{}_{ei}\Omega^{c}{}_{a}{}^{i}\prt_{c}F_{b}{}^{d}+2R_{c}{}^{e}{}_{ei}\Omega^{c}{}_{a}{}^{i}\prt_{d}F_{b}{}^{d}-R_{aebi}\Omega^{dci}\prt_{d}F_{c}{}^{e}+R_{aebi}\Omega_{c}{}^{ci}\prt_{d}F^{de}\nonumber\\
&&+R_{bice}\Omega^{c}{}_{a}{}^{i}\prt_{d}F^{de}-2R_{bcdi}\Omega^{dci}\prt_{e}F_{a}{}^{e}-R_{adbi}\Omega^{dci}\prt_{e}F_{c}{}^{e}-R_{cedi}\Omega^{dci}\prt^{e}F_{ab}\nonumber\\
&&+3R_{bedi}\Omega^{dci}\prt^{e}F_{ac}-R_{bide}\Omega^{dci}\prt^{e}F_{ac}-2R_{bedi}\Omega_{c}{}^{ci}\prt^{e}F_{a}{}^{d}+R_{bide}\Omega_{c}{}^{ci}\prt^{e}F_{a}{}^{d}\nonumber\\
&&-2R_{cedi}\Omega^{c}{}_{a}{}^{i}\prt^{e}F_{b}{}^{d}+R_{bedi}\Omega^{c}{}_{a}{}^{i}\prt^{e}F_{c}{}^{d}-2R_{bide}\Omega^{c}{}_{a}{}^{i}\prt^{e}F_{c}{}^{d}
\bigg]
\label{hcoupling}
\eeqa
while one B-field, one gauge field and two scalar field couplings could be derived as follows:
\beqa
S_{Bf\phi\phi}&\!\!\!\!\!=\!\!\!\!\!&-\frac{\pi^2\alpha'^2T_p}{48}\frac{1}{2}\int d^{p+1}x\sqrt{-\tG}
\bigg[H_{cij}\Om^{bai}\Om^{c}{}_{a}{}^{j}\prt_{d}F_{b}{}^{d}+\frac{1}{2}H_{dij}\Om^{bai}\Om^{c}{}_{a}{}^{j}\prt^{d}F_{bc}
\nonumber\\
&&-H_{dij}\Om^{bai}\Om^{dcj}\prt_{b}F_{ac}-3H_{cij}\Om_{a}{}^{ai}\Om^{cbj}\prt_{d}F_{b}{}^{d}+3H_{cde}\Om_{bai}\Om^{bai}\prt^{e}F^{cd}\nonumber\\
&&+5F^{ab}\Om^{c}{}_{a}{}^{i}\Om_{d}{}^{dj}\prt_{b}H_{cij}+F^{ab}\Om^{c}{}_{a}{}^{i}\Om^{d}{}_{c}{}^{j}\prt_{b}H_{dij}-5F^{ab}\Om^{c}{}_{a}{}^{i}\Om_{d}{}^{dj}\prt_{c}H_{bij}\nonumber\\
&&-3F^{ab}\Om^{c}{}_{a}{}^{i}\Om^{d}{}_{c}{}^{j}\prt_{d}H_{bij}-5H_{cde}\Om_{a}{}^{ai}\Om^{cb}{}_{i}\prt_{b}F^{de}-5H_{dij}\Om^{bai}\Om^{c}{}_{a}{}^{j}\prt_{c}F_{b}{}^{d}\nonumber\\
&&+3F^{ab}\Om^{c}{}_{a}{}^{i}\Om^{ed}{}_{i}\prt_{e}H_{bcd}-F^{ab}\Om^{c}{}_{a}{}^{i}\Om_{d}{}^{d}{}_{i}\prt_{e}H_{bc}{}^{e}+F^{ab}\Om^{c}{}_{a}{}^{i}\Om^{d}{}_{ci}\prt_{e}H_{bd}{}^{e}\nonumber\\
&&+5F^{ab}\Om^{c}{}_{a}{}^{i}\Om_{d}{}^{dj}\prt_{i}H_{bcj}+F^{ab}\Om^{c}{}_{a}{}^{i}\Om^{d}{}_{c}{}^{j}\prt_{i}H_{bdj}+F^{ab}\Om_{c}{}^{ci}\Om_{d}{}^{dj}\prt_{j}H_{abi}\nonumber\\
&&-F^{ab}\Om_{dc}{}^{j}\Om^{dci}\prt_{j}H_{abi}-5F^{ab}\Om^{c}{}_{a}{}^{i}\Om_{d}{}^{dj}\prt_{j}H_{bci}-F^{ab}\Om^{c}{}_{a}{}^{i}\Om^{d}{}_{c}{}^{j}\prt_{j}H_{bdi}\nonumber\\
&&+\frac{1}{2}F^{ab}\Om_{c}{}^{ci}\Om_{d}{}^{d}{}_{i}\prt_{e}H_{ab}{}^{e}-\frac{1}{2}F^{ab}\Om_{dci}\Om^{dci}\prt_{e}H_{ab}{}^{e}-\frac{1}{2}H_{bde}\Om^{bai}\Om^{dc}{}_{i}\prt_{c}F_{a}{}^{e}\bigg]
\label{bfcoupling}
\eeqa
In addition,  one B-field, one gauge field and two scalar field couplings are:
\beqa
S_{Bf\phi\phi}&\!\!\!\!\!=\!\!\!\!\!&-\frac{\pi^2\alpha'^2T_p}{48}\frac{1}{2}\int d^{p+1}x\sqrt{-\tG}
\bigg[F^{ab}\prt_{d}F_{c}{}^{e}\prt^{d}F_{a}{}^{c}\prt_{f}H_{be}{}^{f}-F^{ab}\prt^{d}F_{a}{}^{c}\prt^{e}F_{cd}\prt_{f}H_{be}{}^{f}\nonumber\\
&&-F^{ab}\prt^{d}F_{a}{}^{c}\prt_{e}H_{bdf}\prt^{f}F_{c}{}^{e}-H_{bdf}\prt^{c}F^{ab}\prt^{e}F_{a}{}^{d}\prt^{f}F_{ce}+F^{ab}\prt^{d}F_{a}{}^{c}\prt_{f}H_{bce}\prt^{f}F_{d}{}^{e}\label{bffcoupling}\\
&&
+\frac{3}{4} H_{bef} \prt^{c}F^{ab} \prt_{d}F^{ef} \prt^{d}F_{ac}- \frac{5}{4} H_{def} \prt_{b}F^{ef} \prt_{c}F_{a}{}^{d} \prt^{c}F^{ab}+\frac{5}{12}H_{cdf}\prt^{c}F^{ab}\prt^{e}F_{a}{}^{d}\prt^{f}F_{be} \nonumber\\
&&-  \frac{1}{4} H_{def} \prt_{a}F^{ab} \prt_{c}F^{ef} \prt^{d}F_{b}{}^{c}
-  \frac{1}{4} H_{cef} \prt_{a}F^{ab} \prt_{d}F^{ef} \prt^{d}F_{b}{}^{c} -\frac{1}{4} H_{cdf} \prt_{a}F^{ab} \prt_{b}F^{cd} \prt_{e}F^{ef}\nonumber\\
&&-\frac{1}{4}F^{ab}\prt_{c}H_{bef}\prt_{d}F^{ef}\prt^{d}F_{a}{}^{c}-\frac{1}{4}F^{ab}\prt_{c}F^{ef}\prt_{d}H_{bef}\prt^{d}F_{a}{}^{c}-F^{ab}\prt_{c}F_{a}{}^{c}\prt_{d}F^{de}\prt_{f}H_{be}{}^{f}\nonumber\\
&&-\frac{1}{2}F^{ab}\prt^{e}F^{cd}\prt_{f}H_{abd}\prt^{f}F_{ce}-\frac{1}{2}F^{ab}\prt^{d}F_{a}{}^{c}\prt_{e}H_{bcf}\prt^{f}F_{d}{}^{e}-\frac{1}{2}F^{ab}\prt_{c}F^{cd}\prt_{f}H_{abe}\prt^{f}F_{d}{}^{e}\nonumber
\bigg]
\eeqa
An outline of the paper is as follows: In the next section, we review the constraints that linear T-duality and S-duality may impose on an effective world volume action. In section 3, we construct all couplings of one NSNS and three NS strings with arbitrary coefficients,  and find the coefficients by requiring the consistency of the couplings with the linear dualities.

\section{T-duality and S-duality constraints }

The S-duality and  T-duality   transformations on massless fields are generally nonlinear. By constraining the effective actions to be invariant under these nonlinear transformations, one  may fix all   couplings of bosonic fields \cite{Green:1997tv}. Although this would be a difficult task; see for instance  \cite{Garousi:2014oya,Liu:2013dna,Robbins:2014ara} for the case of nonlinear T-duality. In this paper, however, we are interested only in the world volume couplings of one massless closed and three massless open strings states at order $\alpha'^2$. 
We will see in the following that  the T-duality transformations on gauge fields and  scalar fields are linear
  while closed string under T-duality  do not transform to the open string.
By applying this fact,
one finds out that the higher derivative couplings of one   closed and three open string states have to be invariant under linear T-duality and S-duality transformations.

The full set of nonlinear T-duality transformations has been reported in \cite{TB,Meessen:1998qm,Bergshoeff:1995as,Bergshoeff:1996ui,Hassan:1999bv}. 
We assume a    background including a constant dilaton $\phi_0$ and a metric that is  flat in all directions except for the killing direction $y$, \ie $y$ direction is a circle with radius $\rho$. In addition, we consider  quantum fields to be small perturbations around the background, \ie $G_{\mu\nu}=\eta_{\mu\nu}+2h_{\mu\nu}$  and $G_{yy}={\rho^2}/{\alpha'}(1+2h_{yy})$ where $\mu,\nu\neq y$. For this background, the T-duality transformations  are $e^{2\tilde{\phi_0}}={\alpha'e^{2 \phi_0}}/{\rho^2}$,$ \tilde{G}_{\mu\nu}=\eta_{\mu\nu}$ and $\tilde{G}_{yy}={\alpha'}/{\rho^2}$ while in linear order the quantum fluctuations are as follows:
\beqa
&&\tilde{\phi}=\phi-\frac{1}{2}h_{yy},\,\tilde{h}_{yy}=-h_{yy},\, \tilde{h}_{\mu y}=B_{\mu y},\, \tilde{B}_{\mu y}=h_{\mu y},\,\tilde{h}_{\mu\nu}=h_{\mu\nu},\,\tilde{B}_{\mu\nu}=B_{\mu\nu}\labell{linear}
\eeqa
Also, supposing the $\chi_y$ as the transverse scalar, the T-duality transformation of the world volume gauge field, while it is along the Killing direction, is equal to $\tilde{A}_y=\chi_y$.
The same is true for  $\tilde{\chi}_y=A_y$. Furthermore, the gauge field and the transverse scalar field are invariant under the T-duality when they are not along the Killing direction.
This study concerns with  applying the above linear T-duality transformations to the quantum fluctuations while imposing the full nonlinear T-duality on the    background. 
The latter needs  the DBI effective action to possess  an overall factor of $e^{-\Phi}\sqrt{-\tilde{G}}$.

As it is explained in \cite{Garousi:2009dj,Jalali:2015xca,Jalali:2016xtv}, the effective couplings, which are invariant under the mentioned linear T-duality, can be found as follows: First, in the static gauge, we will introduce, in terms of the world volume indices $a,b,\cdots$ and the transverse indices $i,j,\cdots$, all couplings on the world volume of D$_p$-brane which consist of one massless closed and two open string states. The couplings include all contractions of one massless closed and two open string states that will be introduced in the next section and pull-back of couplings in Eq. \ref{DBIj} on the world volume and transverse space.
This action will be called $S_p$, which then will be reduced to an action in 9-dimensional space. This process  generates two different actions. In one of these actions,  the Killing direction $y$ is a world volume direction, \ie $a=(\tilde{a},y)$,  which we name as $S_p^w$, while in the other one the Killing direction $y$ is a transverse direction, $i=(\tilde{i},y)$ named as $S_p^t$. Up to some total derivative terms, the $S_p^w$  transformation under the linear T-duality  Eq. \reef{linear}, named as $S_{p-1}^{wT}$, should be equal to $S_{p-1}^t$, \ie
\beqa
S_{p-1}^{wT}-S_{p-1}^t&=&0\labell{Tconstraint}
\eeqa
The unknown coefficients in the original action $S_p$ will be constrained by this requirement.

On the other hand, the invariance  of type IIB theory under S-duality transformations produces another set of constraints on the coefficients of $S_p$.  Due to S-duality, the  graviton in  Einstein frame, \ie  $G^E_{\mu\nu}=e^{-\Phi/2}G_{\mu\nu}$, and the transverse scalar fields are  invariant. Also, the following fields will transform as doublets \cite{ Gibbons:1995ap,Tseytlin:1996it,Green:1996qg}:
\beqa\label{sdtrans}
\cB\ \equiv\ 
\pmatrix{B \cr 
C^{(2)}}\rightarrow (\Lambda^{-1})^T \pmatrix{B \cr 
C^{(2)}}\,\,\,\\
\cF\ \equiv\ \pmatrix{*F \cr 
G(F) }\rightarrow (\Lambda^{-1})^T \pmatrix{*F \cr 
G(F) }\nonumber
\eeqa
where the matrix $\Lambda \in SL(2,Z)$  while $G(F)$ is a nonlinear function of $F,\, \Phi,\, C$. Also in the last equation  we have   $(*F)_{ab}=\epsilon_{abcd}F^{cd}/2$. As reported by \cite{ Gibbons:1995ap}, to the linear order of the quantum fluctuations and nonlinear background, which we call  linear S-duality, one has   $G(F)=e^{-\phi_0}F$ where $\phi_0$ is the constant dilaton background.    The transformation of the dilaton and the RR scalar, where the latter will be called $C$, appears in  $SL(2,Z)$ transformation of   the  matrix $\cM$ 
\beqa
 {\cal M}=e^{\Phi}\pmatrix{|\tau|^2\ C \cr 
C\ 1}\label{M}
\eeqa
in which  $\tau=C+ie^{-\Phi}$. This matrix  transformation is as follows \cite{ Gibbons:1995ap}:
\beqa
{\cal M}\rightarrow \Lambda {\cal M}\Lambda ^T\labell{TM}
\eeqa
To the zero order  of quantum fluctuations  and nonlinear order of the background field $\phi_0$, the matrix $\cM$ has been found  as
\beqa
 \cM_0=\pmatrix{e^{-\phi_0}& 0\cr 
0&e^{\phi_0}},
\eeqa
while in  the first order  we have
\beqa
\delta\cM=\pmatrix{-e^{-\phi_0}\phi& e^{\phi_0}C\cr 
 e^{\phi_0}C&e^{\phi_0} \phi}.
\eeqa
The behavior of the above two matrices under the $SL(2,Z)$ transformation is as \reef{TM}.

Using the above transformations, one can easily show that there won't be any couplings between one dilaton and two transverse scalars in the Einstein frame. 
It could be found from  $\Tr(\cM_0^{-1}\delta\cM)=0$ that one can not be able to make a $SL(2,Z)$ invariant combination of $\cM_0$ and $\delta\cM$.
 This requirement generates a set of constraints on the coefficients of the effective action S$_p$.

One can easily find the B-fields and three gauge field couplings, appearing  in the S-dual multiplet, as follow:
\beqa 
\left(\prt (*\cF^T)\cM_0\prt^2\cF\right)\prt \left((*\cF^T)\cM_0\prt^2\cB\right)=e^{-2\phi_0}\left(\prt(*F)\prt(*F)+\prt F\prt F\right)
\left(\prt F\prt^2 B+\cdots \right)\label{bsdual}
\eeqa
In the above equation the dots  refers to the $C^{(2)}$ RR antisymmetric two form couplings which  we do not consider in this work.
Furthermore, it is possible to show the invariance of the following structures  under the linear S-duality transformation.
\beqa 
R\,\Om\,\cF^T \cM_0\prt\cF &=& e^{-\phi_0}R\,\Om\left(*F\prt(*F)+ F\prt F\right) \labell{doub}\\\nonumber
\Om\,\prt\cF^T\prt\cM\prt\cF&=& \Om\left(e^{-\phi_0}\prt\Phi\prt F\prt F-e^{-\phi_0}\prt\Phi\prt (*F) \prt (*F)
+\cdots\right)
\eeqa
 Then the couplings of one closed and three  open string states on the world volume of D$_3$-brane should appear, up to total derivative terms, in the above structures.
These structures  constrain some of the coefficients of the couplings in the effective action $S_p$.

Following the discussion in \cite{Robbins:2014ara}, we know that for the probe branes action to be constructed,  we need to impose   the bulk equations of motion  at order $\alpha'^0$ into $S_p$. Because the world volume couplings which have linear closed string fields are our main interest,  the supergravity equations of motion at linear order should be imposed, \ie
\beqa
R+4\nabla^2\Phi&=&0\nonumber\\
R_{\mu\nu}+2\nabla_{\mu\nu}\Phi&=&0\nonumber\\
\nabla^{\rho}H_{\rho\mu\nu}&=&0
\eeqa
in which $\mu,\nu,\rho$ show the bulk indices.
These indices could be rewritten based on world volume and transverse indices as follows
\beqa
R_{\mu\hph{i}\nu i}^{\hph{a}i} &=&  -2\nabla_{\mu\nu}\Phi-R_{\mu\hph{c}\nu c}^{c}\nonumber\\
\nabla^i{}_i\Phi&=&-\nabla^a{}_a\Phi\nonumber\\
\nabla^{i}H_{i\mu\nu}&=&-\nabla^{a}H_{a\mu\nu}\labell{eom}
\eeqa
This illustrate that the terms on the lhs are not    independent. Therefore, the coefficients of the couplings in the effective action $S_p$, which involve the terms on the left-hand side of the  above equation, have to be zero. 

\section{All contractions}
By using ``xAct'' \cite{CS}, a  mathematica  package, all couplings of one massless NSNS  state and three massless NS state   with unknown coefficients will be derive in this section. Then, we constrain the coefficients by imposing the consistency of the couplings with both linear T-duality and S-duality. 

The couplings of one graviton, two gauge fields and   one scalar field  have two structures \ie $R\prt F\Omega F$ and $\Omega\Omega\prt F\prt F$. Each structure has two contractions; the first  one is the following:
\beqa
S^{(1)}_{h\phi ff}&=&-\frac{\pi^2\alpha'^2T_{p}}{48}\int d^{p+1}x\,e^{-\Phi}\sqrt{-\tG}\bigg(\eta_1 R_{abci}\prt_d F^{bc}\Omega_e{}^{ei}F^{ad}+\eta_2 R_{bide}\partial_{a}F^{de} \Omega_{c}{}^{ci} F^{ab}\nonumber\\
&&   \qquad \qquad \qquad+\eta_3R_{d}{}^{e}{}_{ei} \partial_{b}F_{a}{}^{d} \Omega_{c}{}^{ci} F^{ab} +\cdots+\eta_{58} R_{bide} \partial^{e}F_{c}{}^{d}\Omega^{c}{}_{a}{}^{i} F^{ab}\bigg)\label{hphff1}
\eeqa
where $\eta_i$ are 58 arbitrary coefficients that should be determined by imposing appropriate constraints. By imposing the Bianchi identities and also ignoring total derivative terms, one realizes  that not all the mentioned coefficients are independent. One may first find independent coefficients and then impose the constraints. Also, it is possible to first impose the constraints and then ignore the terms that are related by the Bianchi identities and total derivatives. We use the latter approach which is easier to work with computer. Graviton also may appear in the second fundamental form; thus, the couplings of one graviton, one scalar field and two gauge fields appears in the  $\Omega\Omega\prt F\prt F$ structure and the second contraction will be found as follow:
\beqa
S^{(2)}_{h\phi ff}&=&-\frac{\pi^2\alpha'^2T_{p}}{48}\int d^{p+1}x\,e^{-\Phi}\sqrt{-\tG}\bigg(\phi_1 \Omega^{bai} \Omega^{c}{}_{ai} \partial_{b}\mathcal{F}^{de} \partial_{c}\mathcal{F}_{de}+\phi_2 \Omega_{a}{}^{ai} \Omega^{cb}{}_{i} \partial_{b}\mathcal{F}^{de} \partial_{c}\mathcal{F}_{de}\nonumber\\
&& \qquad  \qquad \qquad \qquad+\phi_3 \Omega^{bai} \Omega^{dc}{}_{i} \partial_{b}\mathcal{F}_{de} \partial_{c}\mathcal{F}_{a}{}^{e}+\cdots+\lambda_8 \Omega_{abi} \Omega^{abi} \Omega_{cdj} \Omega^{cdj}\bigg)\label{hphff2}
\eeqa
in which $\cF_{ab}=F_{ab}+B_{ab}$ is the gauge invariant combination of B-filed and gauge field open string. Through the $\cF_{ab}$ definition the couplings in \reef{hphff2}  include either one graviton, one scalar and two gauge fields or one B-filed, one gauge field and two scalar field couplings. It should be pointed out that $\eta_i$ and $\phi_i$ are 66 arbitrary coefficients.

To check the consistency of the couplings in \reef{hphff1} and  \reef{hphff2}  with T-duality, we need some couplings of dilaton and B-field. The dilaton couplings are as follow:
\beqa
S^{(1)}_{\Phi\phi ff}&=&-\frac{\pi^2\alpha'^2T_{p}}{48}\int d^{p+1}x\,e^{-\Phi}\sqrt{-\tG}\bigg(\zeta_1 \Omega^{ab}{}_{i} \partial_{a}F^{cd} \partial_{b}F_{cd} \partial^{i}\Phi + \zeta_2 \Omega^{ab}{}_{i} \partial_{c}F_{a}{}^{c} \partial_{d}F_{b}{}^{d} \partial^{i}\Phi \nonumber\\
&&\qquad\qquad\qquad\qquad+ \zeta_3 \Omega^{a}{}_{ai} \partial_{b}F^{bc} \partial_{d}F_{c}{}^{d} \partial^{i}\Phi +\cdots+ \zeta_9 \Omega^{a}{}_{ai} \partial_{d}F_{bc} \partial^{d}F^{bc} \partial^{i}\Phi\label{dilphff1}
\bigg)
\eeqa
\beqa
S^{(2)}_{\Phi\phi ff}&=&-\frac{\pi^2\alpha'^2T_{p}}{48}\int d^{p+1}x\,e^{-\Phi}\sqrt{-\tG}\bigg( \chi_1 F^{ab} \Omega^{c}{}_{c}{}^{i} \partial_{d}F_{b}{}^{d} \partial_{i}\partial_{a}\Phi -  \chi_2 F^{ab} \Omega_{a}{}^{ci} \partial_{d}F_{c}{}^{d} \partial_{i}\partial_{b}\Phi \nonumber\\
&&\qquad\qquad\qquad\qquad+\chi_3 F^{ab} \Omega^{cdi} \partial_{d}F_{bc} \partial_{i}\partial_{a}\Phi+\cdots + \chi_{11} F^{ab} \Omega^{cdi} \partial_{d}F_{ab} \partial_{i}\partial_{c}\Phi\label{dilphff2} 
\bigg)
\eeqa
where $\zeta_i$ and $\chi_i$ are 20 arbitrary coefficients and $\Phi$ stands for the dilaton.
Furthermore, the couplings of one graviton and three scalar fields in addition to the pull-back of couplings in the \reef{DBIj}, which was mentioned in the introduction, can appear only through the second fundamental form as follow:
\beqa
S_{h\phi\phi\phi}&=&-\frac{\pi^2\alpha'^2T_{p}}{48}\int d^{p+1}x\,\bigg(\lambda_1 \Omega^{a}{}_{a}{}^{i} \Omega^{b}{}_{bi} \Omega^{c}{}_{c}{}^{j} \Omega^{d}{}_{dj}+\lambda_2 \Omega^{a}{}_{a}{}^{i} \Omega^{b}{}_{b}{}^{j} \Omega_{cdj} \Omega^{cd}{}_{i}\nonumber\\
&& \qquad  \qquad \qquad \qquad+\lambda_3 \Omega_{a}{}^{c}{}_{i} \Omega^{abi} \Omega_{b}{}^{dj} \Omega_{cdj}+\cdots+\lambda_8 \Omega_{abi} \Omega^{abi} \Omega_{cdj} \Omega^{cdj}\bigg)\label{hphphph}
\eeqa
in which $\lambda_i$ are 8 unknown coefficients that should be determined by imposing appropriate constraints.
To assure the T-duality of one graviton and three scalar fields, we consider one dilaton and three scalar field couplings as follow:
\beqa
S_{\Phi\phi\phi\phi}&=&-\frac{\pi^2\alpha'^2T_{p}}{48}\int d^{p+1}x\,e^{-\Phi}\sqrt{-\tG}\bigg( \rho_1 \Omega^{a}{}_{ai} \Omega^{b}{}_{b}{}^{j} \Omega^{c}{}_{cj} \partial^{i}\Phi + \rho_2 \Omega^{a}{}_{a}{}^{j} \Omega_{bcj} \Omega^{bc}{}_{i} \partial^{i}\Phi\nonumber\\
&&\qquad  \qquad \qquad \qquad +\rho_3 \Omega_{a}{}^{cj} \Omega^{ab}{}_{i} \Omega_{bcj} \partial^{i}\Phi+ \rho_4 \Omega^{a}{}_{ai} \Omega_{bcj} \Omega^{bcj} \partial^{i}\Phi \bigg)\label{dilphphph}
\eeqa
Also, the couplings of one  B-filed, one gauge field and two scalar fields appear in two structures \ie $H\prt F\Omega\Omega$ and $\prt H F\Omega\Omega$. We consider these couplings as follow:
\beqa
S^{(1)}_{B\phi\phi f}&=&-\frac{\pi^2\alpha'^2T_{p}}{48}\int d^{p+1}x\,e^{-\Phi}\sqrt{-\tG}\bigg(\iota_1 H_{cde} \Omega^{bai} \Omega^{c}{}_{ai} \partial_{b}F^{de} + \iota_2 H_{dij} \Omega^{bai} \Omega^{c}{}_{a}{}^{j} \partial_{c}F_{b}{}^{d}\nonumber\\
&&\qquad  \qquad \qquad \qquad+ \iota_3 H_{cde} \Omega_{a}{}^{ai} \Omega^{cb}{}_{i} \partial_{b}F^{de} +\cdots+\iota_{14} H_{cde} \Omega_{bai} \Omega^{bai} \partial^{e}F^{cd}\bigg)\nonumber\\
S^{(2)}_{B\phi\phi f}&=&-\frac{\pi^2\alpha'^2T_{p}}{48}\int d^{p+1}x\,e^{-\Phi}\sqrt{-\tG}\bigg( \psi_1 F^{ab} \Omega_{cb}{}^{j} \Omega^{c}{}_{a}{}^{i} \partial_{d}H^{d}{}_{ij}- \psi_2 F^{ab} \Omega^{c}{}_{a}{}^{i} \Omega_{d}{}^{dj} \partial_{b}H_{cij}\nonumber\\
&& \qquad  \qquad \qquad \qquad+ \psi_3 F^{ab} \Omega^{c}{}_{a}{}^{i} \Omega_{d}{}^{dj} \partial_{c}H_{bij}+\cdots+\psi_{25} F^{ab} \Omega^{c}{}_{a}{}^{i} \Omega^{d}{}_{b}{}^{j} \partial_{j}H_{cdi} \bigg)\label{bphphf2}
\eeqa
where $\iota_i$ and $\alpha_i$ are 39 arbitrary coefficients that should be determined by imposing appropriate constraints. As explained before, some of the couplings of one B-field, one gauge field and two scalar fields have been included in Eq. \reef{hphff2}. Finally, the couplings of one B-field  and three gauge fields can be introduced:
\beqa
S^{(1)}_{Bf f f}&\!\!\!\!=\!\!\!\!&-\frac{\pi^2\alpha'^2T_{p}}{48}\int d^{p+1}x\,e^{-\Phi}\sqrt{-\tG}\bigg(\tau_1 H_{def} \partial_{a}F^{ab} \partial_{b}F^{cd} \partial_{c}F^{ef} + \tau_2  H_{def} \partial_{a}F_{c}{}^{d} \partial_{b}F^{ef} \partial^{c}F^{ab} \nonumber\\
&&\qquad   \qquad \qquad+ \tau_3  H_{def} \partial_{b}F^{ef} \partial_{c}F_{a}{}^{d} \partial^{c}F^{ab} +\cdots+\tau_{29}H_{def} \partial_{c}F_{ab} \partial^{c}F^{ab} \partial^{f}F^{de}\bigg)\nonumber\\
S^{(2)}_{Bf f f}&\!\!\!\!=\!\!\!\!&-\frac{\pi^2\alpha'^2T_{p}}{48}\int d^{p+1}x\,e^{-\Phi}\sqrt{-\tG}\bigg( \gamma_1  \partial_{b}H_ {def} F^{ab} \partial_{a}F^{cd} \partial_{c}F^{ef} + \gamma_2 \partial_{b}H_ {aef}F^{ab}  \partial_{c}F^{cd} \partial_{d}F^{ef} \nonumber\\
&& \qquad   \qquad \qquad+ \gamma_3 \partial_{d}H_{bef}F^{ab} \partial_{a}F^{cd} \partial_{c}F^{ef} +\cdots+\gamma_{81}  \partial_{f}H_{cde} F^{ab} \partial^{c}F_{ab} \partial^{f}F^{de} \bigg)\nonumber\\
S^{(3)}_{Bf f f}&\!\!\!\!=\!\!\!\!&-\frac{\pi^2\alpha'^2T_{p}}{48}\int d^{p+1}x\,e^{-\Phi}\sqrt{-\tG}\bigg( \delta_1 \partial_{c}\mathcal{F}^{ef} \partial^{c}\mathcal{F}^{ab} \partial_{d}\mathcal{F}_{ef} \partial^{d}\mathcal{F}_{ab} + \delta_2 \partial_{a}\mathcal{F}^{ab} \partial_{c}\mathcal{F}^{ef} \partial_{d}\mathcal{F}_{ef} \partial^{d}\mathcal{F}_{b}{}^{c} \nonumber\\
&&   \qquad \qquad \qquad+\delta_3 \partial_{b}\mathcal{F}_{ef} \partial^{c}\mathcal{F}^{ab} \partial_{d}\mathcal{F}_{c}{}^{f} \partial^{e}\mathcal{F}_{a}{}^{d} +\cdots+\delta_{33} \partial_{c}\mathcal{F}_{ab} \partial^{c}\mathcal{F}^{ab} \partial_{f}\mathcal{F}_{de} \partial^{f}\mathcal{F}^{de}\bigg)\label{bfff1}
\eeqa
in which $\tau_i$, $\gamma_i$ and $\delta_i$ are 143 arbitrary coefficients that should be deduced by imposing linear dualities.

Now we consider the sum of the couplings in \reef{DBIj}, \reef{hphff1}, \reef{hphff2}, \reef{dilphff1}, \reef{dilphff2}, \reef{hphphph}, \reef{dilphphph}, \reef{bphphf2}, \reef{bfff1} \ie
\beqa
S_p&=&S^{(1)}_{h\phi ff}+S^{(2)}_{h\phi ff}+S^{(1)}_{\Phi\phi ff}+S^{(2)}_{\Phi\phi ff}+S_{h\phi\phi\phi}+S_{\Phi\phi\phi\phi}\nonumber\\
&&\qquad\quad\!\,\!\!+S^{(1)}_{B\phi\phi f}+S^{(2)}_{B\phi\phi f}+S^{(1)}_{B\phi f f}+S^{(2)}_{B\phi f f}+S^{(3)}_{B\phi f f}+S_{p.l}\label{sp1}
\eeqa

as the effective action, in $\alpha^{\prime~2}$ order of one closed and three open  strings, and apply the T-duality constraint \reef{Tconstraint}. This procedure gives the following relations between the  constants:
\beqa
&&\iota_4=0,\ \iota_{18}=-1/2,\ \eta_{37}=1+\eta_{13}+\eta_{17}-\eta_{21}-\eta_{22}-\eta_{28}-\eta_{29}-\eta_{36}+2\eta_{19} \nonumber\\
& &\ga_{33}=1-\ga_{13}-\ga_{14}-\ga_{32},\ \ga_{35}=-1+\ga_{13}+\ga_{14}+\ga_{28}-\ga_{29}+\ga_{30}+\ga_{32}+2\ga_{34}\ \nonumber\\
&&\ga_{65}=-1+\ga_{28}-\ga_{29}+\ga_{62}-\ga_{63}+\ga_{64},\ \ga_7=-3/4-\ga_{11}/2+\ga_{14}/2\nonumber\\
&&-\ga_{22}+\ga_{23}/2-\ga_{25}/2+\ga_{26}/2-\ga_{27}/2-\ga_{29}/2-\ga_{3}
-\ga_{4}+\ga_{40}/2+\ga_{41}+\ga_{42}/2\nonumber\\
&&+\ga_{44}+\ga_{47}/2-\ga_{49}/2+\ga_{59}/2-\ga_{6}
+\ga_{60}/2+\ga_{62}/2-\ga_{63}/2+\ga_{67}/2+\ga_{68}/2\nonumber\\
& &
\eta_{39}=-2-\eta_{13},\ \eta_{5}=1-4\ga_1-4\ga_{10}+2\ga_{11}+2\ga_{12}-2\ga_{13}-4\ga_{14}-2\ga_{26}+2\ga_{27}
 \nonumber\\
&&-2\ga_{28} +2\ga_{29}+4\ga_{4}+2\ga_{40}-2\ga_{42}+4\ga_{43}-4\ga_{44}-2\ga_{47}+2\ga_{49}
+2\ga_{58}-2\ga_{60}
-2\ga_{62} \nonumber\\
&&+2\ga_{63}-\eta_{17}+\eta_{26} ,\ \iota_8=-1/4-\iota_{12}/2\ga_{17}=1/2-\ga_{15}-\ga_{16},\ \ga_{21}=\ga_{19}/2-\ga_{20}/2 \nonumber\\\label{tdconst} &&\ga_{35}=-1+\ga_{13}+\ga_{14}-\ga_{28}-\ga_{29}+\ga_{30}+\ga_{32}+2\ga_{34},\ \cdots
\eeqa
where the dots, here and in the following, refers to the constraints which do not include any constant number and only proviode a relationship between unknown coefficients.

As can be seen,  all coefficients of the effctive action, \ie  $S_p$, in the \reef{sp1} are not fixed by imposing consistency of the couplings with the linear T-duality. Thus we need further limitations to constrain the effective action. These constraints might be chosen as the imposition of the consistency with S-duality. 
 For example, by imposing S-duality, it is possible to see that  one dilaton and three scalar fields in \reef{dilphphph} must be vanished in the  world volume  of D$_3$-brane in the Einstein frame. This produces the following constraint: 
\beqa
\delta_{25}=1/2-\delta_{16}/2-\delta_{17}/2-\delta_{18}-\delta_{19}/2-\delta_{23}\label{sd1const}
\eeqa
In addition, S-duality constrains  the couplings of one graviton, one scalar field and two gauge fields in  \reef{hphff1} and \reef{hphff2}. These couplings in the world volume of   D$_3$-brane action must appear in the S-duality invariant structures   $R\,\Om\,\prt\cF^T\cM_0\cF=e^{-\phi_0}R\,\Om\,(\prt (*F) (*F)+\prt F  F )$. This condition provides the following constraints: 
\beqa
\delta_{8}&=&-2+4\delta_{1}-\delta_{15}+\delta_{18}+\delta_{19}-2\delta_{21}+2\delta_{25}-\delta_{27}-\delta_{7},\label{sdconst2} \\ \ga_{63}\!\!&=&-1+2\ga_1+2\ga_{10}-\ga_{11}-\ga_{12}+\ga_{13}+2\ga_{14}+\ga_{26}-\ga_{27}+\ga_{28}-\ga_{29}
-2\ga_4-\ga_{40}\nonumber\\
&+&\ga_{42}-2\ga_{43}+2\ga_{44}+\ga_{47}-\ga_{58}+\ga_{60}+\ga_{62},\ \cdots\nonumber
\eeqa
By applying T-duality constraint in \reef{tdconst} and S-duality constraint in \reef{sd1const} and \reef{sdconst2} to the one dilaton, one scalar and two gauge fields couplings  in \reef{dilphff1} and \reef{dilphff2} we found that these  couplings  satisfy the  S-dual structure, \ie the second equation in  \reef{doub}, without any constraint in the coefficient.

As the final S-duality constrain, we check  the consistency  of the B-fields couplings with linear S-duality according to \reef{bsdual}, \ie the S-dual  multiplet, and find the following constraint:
\beqar
\delta_8&=&-2+4\delta_{1}-\delta_{15}+\delta_{18}+\delta_{19}-2\delta_{21}
+2\delta_{25}-\delta_{27}-\delta_{7}\nonumber\\
\ga_{63}\!&=&7/4+2\ga_1+2\ga_{10}-\ga_{11}-\ga_{12}+2\ga_{13}+3\ga_{14}+\ga_{26}-\ga_{27}+2\ga_{28}
\nonumber\\
&-&4\ga_{29}-2\ga_{4}-2\ga_{42}-4\ga_{43}+2\ga_{44}+4\ga_{47}-\ga_{49}-\ga_{58}
+\ga_{60}+\ga_{62},\ \cdots\label{sdconst3}
\eeqar

Now we impose all constraints in \reef{tdconst}, \reef{sd1const}  and \reef{sdconst2}, resulted  from T-duality and S-duality, on the DBI effective action   \reef{sp1} and obtain the T-dual and S-dual invariance effective action.

 On the other hand the  Riemann curvature satisfies the cyclic symmetry while the field strengths satisfy the Bianchi identities. Therefore, these symmetries have to be imposed in   $S^{DBI}_p$. To perform this step, we write all field strengths in  terms of their corresponding  potentials and write the Riemann curvature in terms of 
\beqa
R_{abcd}=\prt_b\prt_ch_{ad}+\prt_a\prt_d h_{bc}-\prt_b\prt_d h_{ac}-\prt_a\prt_c h_{bd}
\eeqa
Then we find that the coefficients    $\ga_i$ with $i=1,2,3,4,5,6,16,18,20,24,26,27,30,32,34,36,\\
37,45,46,53,54,55,56,58,59,60
,61,62,64,66,67,68,83,74,78,76,77,78,80$ and $\eta_i$ with $i=1,2,3,7,11,12,19,24,26,28,29,30,31,33,36,41,44,47,50,53,55,56$ and $\delta_i$ with $i=2,3,7,\\10,
12,13,14,19,20,22,24,30,31$ and $\zeta_i$ with $i=1,5,8$ and   $\tau_i$ with $i=1,2,4,5,10,11,12,13,\\14,15,
16,18,19,21,22,23,24,25,26,27,28,29$ and $\phi_i$ with $i=1,13,16,21,22$ and $\chi_i$ with $i=5,7,10$ and $\psi_i$ with $i=10,11,12,15,16$ disappear from the action. Consequently, the couplings with the above coefficients display  only the cyclic symmetry and the Bianchi identities.   Thus, we ignore such terms in the DBI effective action.  Finally, we find that the couplings with coefficients   $\ga_i$ with $i=10,11,12,13,14,15,19,22,23,25,28,29,38,39,40,41,42,43,44,\\
47,48,49$ and  $\eta_i$ with $i=13,17,21,22$  and $\zeta_i$ with $i=2,3$  and $\iota_i$ with $i=1,11,12$are total derivative terms, so they can be eliminated from the DBI effective action too. 

By Imposing T-duality and S-duality constraints plus Bianchi identity and cyclic symmetry while neglecting the total derivative couplings from the effective action, we find that all unknown coefficients in \reef{sp1} will be fixed except 
  $\delta_{i}$ with $i=1, 6,15,16,17,18,21,23,27,29,32$. 
 
 One may fix the above coefficients by comparing the couplings $S_{h\phi\phi\phi }$ in \reef{hphphph},   $S^{(3)}_{Bf ff}$ in \reef{bfff1} and $S^{(2)}_{h\phi ff}$ in \reef{hphff2}  with the couplings in \reef{fourph}, \reef{fourf} and \reef{twoftwoph} respectively after applying all mentioned constraints. By comparing \reef{hphphph} with \reef{fourph} we find the following relations:
 $$\delta_{15}=1,\ \delta_{29}=-2(2+\delta_{32}-\delta_6)$$
 while comparing \reef{bfff1} with \reef{fourf} yields the the following constraints:
 $$ \delta_{1}= \delta_{6}= \delta_{16}= \delta_{17}= \delta_{18}= \delta_{21}=0,\  \delta_{23}=2,\  \delta_{32}=-2$$
 and comparing \reef{hphff2} with \reef{twoftwoph} provides the following constraints: 
   $$ \delta_{27}=-4$$
To conclude, we point out that by imposing all above constraints to the effective action of \reef{sp1}, we could find all arbitrary coefficients and add them other couplings of one NSNS closed string and three NS open string correction to the DBI action at the $\alpha'^2$ order. As the result, the dilaton couplings can be found as follow:
\beqa
S_{\Phi\phi\phi\phi}&\!\!\!\!\!=\!\!\!\!\!&-\frac{\pi^2\alpha'^2T_p}{48}\int d^{p+1}x\sqrt{-\tG_{ab}}
\bigg[F^{ab}\Om_{d}{}^{d}{}_{i}\prt_{c}F_{ab}\prt^{i}\prt^{c}\Phi+F^{ab}\Om^{d}{}_{ai}\prt_{c}F_{bd}\prt^{i}\prt^{c}\Phi\\
&&-F^{ab}\Om^{d}{}_{ci}\prt_{d}F_{ab}\prt^{i}\prt^{c}\Phi+F^{ab}\Om^{d}{}_{ai}\prt_{d}F_{bc}\prt^{i}\prt^{c}\Phi-2F^{ab}\Om_{cai}\prt_{d}F_{b}{}^{d}\prt^{i}\prt^{c}\Phi\nonumber\\
&&+\frac{1}{2} \Om_{a}{}^{a}{}_{i} \Om_{b}{}^{bj} \Om_{d}{}^{d}{}_{j} \prt^{i}\Phi - 5 \Om_{ba}{}^{j} \Om^{ba}{}_{i} \Om_{d}{}^{d}{}_{j} \prt^{i}\Phi + \Om^{ba}{}_{i} \Om_{dbj} \Om^{d}{}_{a}{}^{j} \prt^{i}\Phi \nonumber\\
&&+\frac{15}{2} \Om_{a}{}^{a}{}_{i} \Om_{dbj} \Om^{dbj} \prt^{i}\Phi+4\Om^{ba}{}_{i}\prt_{b}F_{a}{}^{c}\
\prt_{d}F_{c}{}^{d}\prt^{i}\Phi+4\
\Om_{a}{}^{a}{}_{i}\prt_{d}F_{bc}\prt^{d}F^{bc}\
\prt^{i}\Phi
\bigg]
\eeqa
while the other couplings are reported formerly in \reef{fourph}, \reef{fourf}, \reef{twoftwoph}, \reef{hcoupling}, \reef{bfcoupling} and \reef{bffcoupling}.

On the other hand, the D-brane effective action at the order $\alpha'^2$  should be consistent with S-matrix elements. It would be interesting to verify above couplings to be consistent with the S-matrix elements of one NSNS and three NS states.

{\bf Acknowledgments}:   We would like to thank M. R. Garousi for very valuable discussions. We also  specially grateful  to H. shenavar  for very helpful conversations.

\bibliographystyle{/Users/Nick/utphys} 
\bibliographystyle{utphys} \bibliography{hyperrefs-final}
\providecommand{\href}[2]{#2}\begingroup\raggedright
\endgroup

\end{document}